# Do Chatbots Walk the Talk of Responsible AI?

Susan Ariel Aaronson[1] and Michael Moreno[2]

## Introduction

In April 2025, sixteen-year-old Adam Raine committed suicide. Over the course of several months, the teen confided his suicidal thoughts to OpenAI's ChatGPT chatbot.[3] ChatGPT is not designed or developed to provide therapy, but it did not respond to Adam's prompts with suggestions that he obtain professional help. Moreover, when Adam expressed concern that his parents would blame themselves if he died, ChatGPT reportedly responded, "That doesn't mean you owe them survival," and offered to help draft his suicide note.[4]

Adam's death was not the only example of chatbot misbehavior. OpenAI claims it doesn't permit ChatGPT "to generate hateful, harassing, violent, or adult content."[5] In July 2025, a reporter documented ChatGPT providing users with detailed instructions for self-mutilation, murder, and satanic rituals.[6] OpenAI has also acknowledged that individuals can misuse its systems.[7] But the company has taken some responsibility. Later that April, OpenAI issued a blog post promising to redesign how ChatGPT responds when a user shows signs of mental or emotional distress."[8]

OpenAI is not the only chatbot developer that has created models that act in an irresponsible manner. Researchers have found that almost every chatbot exhibits problems, inaccuracies, sycophantic behavior, and lies.[9] These bots can also respond to prompts with deceptive behavior to achieve a particular goal.[10] Despite these problems, most chatbot developers insist that they are committed to designing, developing, and deploying AI in a responsible manner.

Unfortunately, AI developers and users have no universal definition for responsible AI which would allow companies to monitor their practicers and help users hold bad actors to account. For example, the OECD defines it in adjectives: human-centered, fair, equitable, inclusive, respectful of human rights and democracy, and designed to contribute  positively to the public good.[11] The Government of Canada sees it as a process;  responsible AI means developing and using AI systems in ways that are ethical, transparent and fair, ensuring they do not cause harm or perpetuate biases.[12] Google also defines it as a process but focuses on human oversight, due diligence, and feedback mechanisms to align with user goals, social responsibility, and international human rights principles.[13] Finally, the authors of a meta-study of responsible AI research define responsible AI as a set of voluntary practices that reveal to users that the AI developer cares about AI's direct and indirect impact on people and the planet.[14]

AI chatbots are widely used globally.[15] If a company allows its bots to act in an irresponsible manner, it is likely to lose market share and stakeholder trust.[16] In this paper, we examine if four prominent AI companies "walk the talk"—if they act responsibly when developing and deploying chatbots. Our sample includes three US companies (Google, OpenAI, and xAI) and one Chinese company (DeepSeek). We compare what these AI developers say on their websites and technical documents and then compare how the chatbots respond to prompts about their responsible AI practices. We then suggest strategies policymakers could use to bolster responsible AI practices. We note that these four chatbots are not necessarily a representative sample, but they provide sufficient variation in their approach to responsible AI.

**Methodology**

Herein, we use a mixed methods approach to compare what AI developing firms say about responsible AI with what they do to ensure that their chatbots act in a responsible manner.[17] To ascertain their commitment to responsible AI we looked to see if they talked about their approach to their users and stakeholders and signaled their commitment by joining a formal responsible AI commitment such as in the Hiroshima Process, EU General Purpose Code of Conduct or Canada Voluntary Code of Conduct on the Responsible Development and Management of Generative AI systems.[18] Table 1 compares the bots.

**Table 1: Simple Chatbot Comparison**

| Firm Name (Chatbot) | Formal Affiliation with Government Code of Responsible AI | Responsible AI Website Commitment | Technical Documentation | Chatbot Evaluation |
|---|---|---|---|---|
| OpenAI (GPT4o) | Yes | Yes | Yes | Yes |
| xAI (Grok 3) | No | EU AI Code of Practice Safety and Security Chapter only | No | Yes |
| Google (Gemini 2.5) | Yes | Yes | Yes | Yes |
| High-Flyer (DeepSeek V3) | No | No | Yes | Yes |

We created a composite definition of responsible AI for this study because there is no internationally accepted definition. To be responsible, a developer must refer to at least one of these keywords including "trustworthy," "responsible," "safe," resilient, reliable, explainable, interpretable, "human rights," "ethics," "ethical," "accountable," "sustainable," "purpose," "purposeful," "human-centered," "fair," "equitable," "inclusive," diverse, "democratic", "open," transparent," "alignment," "democracy," "safety," "privacy," "oversight," "public good," and "public interest."

We next use three different methods and venues of analysis.

a. **Websites**: AI developers (companies, governments, and researchers) use websites to communicate to a broad and diverse audience.[19] We searched these websites to answer 4 sub-questions.
   1. Did the firm define and discuss responsible AI in general?
   2. Did the firm discuss how their approach to responsible AI shapes their approach to chatbot design, development, and deployment?
   3. Does the company use any, some, or most of the keywords above to describe their practices?

4. Does the company provide details of who was accountable for irresponsible behavior and how it remedied it?

**b. Technical Documentation:** We next examined whether the company is transparent and accountable regarding model development. We examine if the company discusses key issues such as human rights, human centered, participatory, in these technical documents, if available. We used qualitative data analysis software, NVivo, to search for keywords to signal the importance of these terms in the design of the chatbot. Our keywords and their stem words[20] include "trustworthy," "responsible," "safe," resilient, reliable, explainable, interpretable, "human rights," "ethics," "ethical," "accountable," "sustainable," "purpose," "purposeful," "human-centered," "fair," "equitable," "inclusive," diverse, "democratic," "open," transparent," "alignment," "democracy," "safety," "privacy," "oversight," "public good," and "public interest." We will use three different methods and venues of analysis.

**c. Chatbot Evaluation**: Next, we look at what chatbots do and say. We asked our four chatbots a standardized set of questions, including:
   o How was the "model" (for example, "Grok 3") trained to ensure that it respects user rights? Provide examples.
   o How was the "model" trained to ensure it respects and promotes democratic values? Provide examples.
   o How was the "model" trained to promote fairness and minimize bias in its outputs? Provide examples.
   o How does user feedback affect the "model" development and deployment to align more with responsible AI? Provide examples.
   o How was "model" trained to be inclusive and equitable? Provide examples.

We note some concerns with our methodology. As the linguist Emily Bender and her colleagues noted, chatbots are stochastic parrots. Although they seem human-like, they cannot read text or answer questions as humans do. These bots are designed to predict the likely next word. So, some might argue that evaluating the chatbots doesn't give you significant information about a company's vision of responsible AI.[21] While we agree that the chatbot is repeating words, we believe the answers still reflect training priorities. Hence, our mixed methods approach will provide the authors with information on how and whether firms operationalize responsible AI in their chatbots.

## Findings

### Website Analysis
The four chatbot developers vary in how they use their websites to define and operationalize their commitment to responsible AI—a critical component for building and sustaining public trust in their technologies. Google used its websites to signal its commitment to a broad vision of responsible AI, while OpenAI focused on a narrower set of responsible practices. In contrast, DeepSeek and xAI do not discuss key components of responsible AI.

*1. Did the firm define and discuss responsible AI in general*?

Google connects responsible AI to the firm's mission "to organize the world's information" and its internal AI principles.[22] Moreover, Google does extensive research into responsible AI[23] and it publishes a yearly responsible AI report.[24] OpenAI and, to a lesser extent, xAI focused on safety, but neither firm describes their efforts as part of their approach to responsible AI.[25]

Both OpenAI and xAI highlight their obligations and responsibilities to humanity, stating: "The mission of OpenAI is to ensure artificial general intelligence (AGI) benefits all of humanity." Thus, safety is core to the company's mission."[26] xAI explicitly states, "We build AI specifically to advance human comprehension and capabilities.[27]

On its US site, DeepSeek says nothing about responsible AI.[28] However, in October, on its Pakistani blog site, the company stated that it would invest in bias mitigation; continue to work in a transparent manner; protect privacy; enhance human capability;  and optimize energy usage."[29] The authors do not know why this commitment does not appear on US websites as a company's responsibility efforts should be global.

2.  *Did the firm discuss how their approach to responsible AI shapes their approach to chatbot design, development, and deployment?*

No company in our sample addressed how their responsible AI commitments shape their approaches to chatbot design, development, and deployment. Both Google and OpenAI are transparent about how they train their models, but neither company directly connects these practices to responsible AI principles.[30]

3.  *Does the company use any, some, or most of the keywords above to describe their practices?*

OpenAI and Google use many of these terms. For example, OpenAI notes, "We teach our AI good behavior so it can be both capable and aligned with human values."[31] In addition, the company notes, "We work to develop AI that elevates humanity and promotes democratic ideals…Decisions about how AI behaves and what it is allowed to do should be determined by… society."[32] Google has devoted several web pages to inform its stakeholders about responsible AI, because it says it has a responsibility to build AI that works for everyone.[33] It defines it as "as a living constitution, keeping us "motivated by a common purpose." The company says it uses tools  and resources such as "Explainable AI, Model Cards, and the TensorFlow open-source toolkit to provide model transparency in a structured, accessible way."[34] The company has also prepared a Responsible Generative AI toolkit.[35]

Google says its approach to responsible AI is grounded in 3 principles: bold innovation, responsible development, and collaborative progress together.[36] The company says it will implement human oversight, due diligence and feedback mechanisms; invest in industry-leading approaches to advance safety and security, employ rigorous design testing, monitoring, and safeguard and promote privacy and security, and respect intellectual property rights. However, none of the firms used their websites to discuss how the firm trains its chatbots to embody these principles.

4.  *Does the company provide details of who was accountable for irresponsible behavior and how it remedied it?*

Accountability is a key concept for responsible AI. However, none of the 4 companies used their web pages to delineate who at the staff, management or board level was accountable for ensuring responsible AI. Without such information, users and policymakers will struggle to hold these firms to account when they seek to inform the developer about problems with a specific chatbot.

### Technical Documentation Analysis

Next, the authors examined how the four firms discussed the design and development of their chatbots in widely available technical documents. Developers utilize these documents to delineate how they developed a model. Three companies in our sample—OpenAI,[37] Google,[38] and DeepSeek[39]—released technical reports for analysis. Grok published technical documents after our research concluded.[40] Hence, this section does not discuss Grok.[41]

The chatbot firms did not appear to integrate responsible AI into their technical reports, although they did utilize some responsible AI terms.[42] For example, OpenAI mentioned the need for "responsible and safe societal adoption"[43] of language models and also emphasized that "warnings and user education documents are essential to responsible uptake of increasingly powerful language models like GPT-4."[44] OpenAI also discussed how it developed mitigation strategies to "reduce the risk that our models are used in a way that could violate a person's privacy rights."[45] The company did not reference other key terms such as "democratic," "human rights," etc.…Rather, the company framed responsible AI primarily through technical safety measures and privacy protections, consistent with their web-based materials.

In contrast, Google utilized a broader swath of responsible AI terms. The company outlined its approach to responsible AI in a technical document section entitled "Safety, Security, and Responsibility."[46] Google stated they are "committed to developing Gemini responsibly, innovating on safety and security alongside capabilities."[47] The company also detailed prohibitive behaviors (such as ensuring that the bot did not encourage violence) as well as positive behaviors (such as providing multiple perspectives when consensus does not exist). Google mentioned "democratic" once, but only in the narrow context of external researchers evaluating "democratic harms and radicalization."[48] Google also noted it established a Google DeepMind Responsibility and Safety Council (RSC), to "review initial ethics and safety assessments on novel model capabilities in order to provide feedback and guidance during model development."[49] Finally, to ensure responsible behavior, the company described comprehensive training methodologies, including automated red teaming and reinforcement learning from human feedback.[50]

In contrast, DeepSeek's technical documentation contained no reference to responsible AI. It mentioned "safety" only once in a chart with minimal context or elaboration.[51] The company's documentation focused exclusively on technical aspects and performance metrics, providing little discussion of responsible AI considerations. However, in September 2025, DeepSeek researchers published a paper examining how the model was trained to tackle reasoning tasks such as coding or mathematical reasoning through a trial and error approach called reinforcement learning.[52] Scientific American reported that DeepSeek was the first LLM to undergo peer review—meaning it made its underlying model and data available to other researchers to replicate.[53] In so doing, the company used its technical document to model responsible AI.

A key element of responsible AI is to ensure that humanity benefits from AI deployment (alignment). However, no one really knows if a particular model is aligned with societal needs. Not surprisingly, the firms took different approaches. OpenAI claimed that by using reinforcement learning, its model would yield responses that align with the users' intent.[54] DeepSeek described how it worked to align its model with human preferences[55] through supervised fine-tuning and reinforcement learning across "diverse domains.[56] In contrast, Google focused on deceptive alignment. The company dedicated a section of its technical report to evaluating the chatbot's stealth and situational awareness capabilities."[57]

We note that technical documents are not designed to explain chatbot responsibility, but firm utilization of these words signal their importance to the firm. Hence, we utilized a word count to better understand the number of words devoted to responsible AI. As Table 2 below illustrates, firms rarely used these terms. They totaled 391 relevant mentions of our keywords out of 97,854 total words, comprising only .004% across the three technical documents. Moreover, some important responsible AI terms were barely mentioned, while other related terms, such as public goods, public interest, human rights, sustainable, human-centered, or equitable were not mentioned at all. Hence, the researchers were unable to ascertain how the firms translated these terms into responsible chatbot behavior.

| | | | | | |
|---|---|---|---|---|---|
| **Technical Documentation Word Frequency Count** | | | | | |
| **Key Words:** | **Total Word Frequency:** | **OpenAI Word Frequency:** | **DeepSeek Word Frequency:** | **Gemini Word Frequency:** | **Related words included:** |
| *Responsible* | 12 | 4 | - | 8 | Responsibility; responsiveness; responsibly |
| *Human Rights* | 0 | - | - | - | - |
| *Ethics* | 4 | 3 | - | 1 | Ethical; ethic |
| *Accountability* | 7 | 7 | - | - | Account; accountable; accounted |
| *Sustainable* | 0 | - | - | - | Sustain; sustainability |
| *Purpose* | 1 | 1 | - | - | Purposeful, repurpose |
| *Safety* | 184 | 108 | 1 | 75 | Safe |
| *Resilient* | 9 | 2 | - | 7 | Resilience |
| *Reliable* | 15 | 11 | 3 | 1 | Reliability; unreliable |
| *Explainable* | 1 | 1 | - | - | Explainability |
| *Interpretable* | 1 | 1 | - | - | Interpret |
| *Human-centered* | 0 | - | - | - | - |
| *Fair* | 14 | 8 | 1 | 5 | Fairly, unfair; fairness |
| *Equitable* | 0 | - | - | - | Equity |
| *Inclusive* | 2 | 2 | - | - | Inclusivity |
| *Diverse* | 18 | 2 | 9 | 7 | Diversity |
| *Democratic* | 2 | 1 | - | 1 | Democracy |
| *Open* | 53 | 9 | 34 | 10 | Openness |
| *Transparent* | 6 | 6 | - | - | Transparency |
| *Alignment* | 42 | 23 | 14 | 5 | Align; aligned; unaligned; misalign |
| *Privacy* | 13 | 8 | - | 5 | - |
| *Oversight* | 7 | 4 | - | 3 | Oversee |
| *Public good* | 0 | - | - | - | - |
| *Public interest* | 0 | - | - | - | - |
| *Total* | **391** | **201** | **65** | **125** | **(of 97,896 total words)** |

**Chatbot Evaluation Analysis**

The researchers developed a set of questions to understand whether and how the bot was trained to discuss issues and principles associated with responsible AI. The researchers catalogued these responses in a spreadsheet available on the Hub's website. Readers should refer to the spreadsheet for full chatbot responses.

1. *How was the "Model" (for example, "Grok 3") Trained to Ensure that it Respects User Rights? Provide Examples.*

When we asked the model's respect for user rights, all four companies focused on privacy. However, the researchers were surprised that the bots did not address any other user rights, such as freedom of expression or the difficulties of balancing various human rights simultaneously. The bot's responses to this prompt varied. ChatGPT-4o stated that "OpenAI trained GPT-4o to avoid using personal, confidential, or proprietary data from private sources." Moreover, the model is designed to forget user interactions after the session ends, unless the user explicitly consents to retain data via custom instructions" (see summary).[58] In contrast, Gemini detailed user control options, and Grok 3 provided opt-out instructions.[59] DeepSeek's V3 responded that if "a user asks for personal data (e.g., 'What's John Doe's phone number?'), DeepSeek v3 denies the request.[60]

2. *How was the "Model" Trained to Promote Fairness and Minimize Bias in its Outputs? Provide Examples.*

When we asked about training to promote fairness and minimize bias, all four bots responded to the prompt stating that they filtered training data to reduce bias. For example, ChatGPT-4o responded that although it was trained on publicly available and licensed data, it didn't use personal, confidential, or proprietary data from private sources without explicit permission. It also claimed that it excluded content from extremist websites or disinformation hubs.[61] DeepSeek V3 made similar claims about data filtering, stating "the training data was carefully curated to avoid copyrighted, sensitive, or personally identifiable information (PII) without proper authorization" and that "Pre-training filtering: Removed overtly biased or hateful content from datasets."[62] Grok 3 responded that it had been trained on a wide range of datasets. at "Biases in training data or user inputs on X can inadvertently influence outputs, and xAI addresses this through ongoing monitoring and refinement."[63] In contrast, Google's Gemini focused on the scope and variety of training data, noting its developers curated "diverse datasets that aim to represent a wide range of cultures, demographics, and perspectives" while noting they employ "techniques to identify and, where possible, filter out or re-balance biased content within the training data."[64]

3. *How was the "Model" Trained to be Inclusive and Equitable? Provide Examples.*

When we asked how the model was trained to be inclusive and equitable, Google Gemini and DeepSeek responded with specific examples. For example, OpenAI explained that "it integrates input from experts and diverse stakeholders, including ethics researchers, civil society, and international human rights guidance.[65] Gemini responded that it created multimodal applications to facilitate inclusion (for example, so blind or deaf people could use the model), and it supported 1000 languages. DeepSeek V3 provided detailed examples such as addressing crime rate disparities through "Historical redlining, underfunded schools, and policing biases—not race itself." The bot also responded that it utilized language guidance, recommending terms like "chairperson" versus "chairman" and "disabled people" versus "the disabled."[66] In contrast, Grok 3 responded that it utilized data across "cultural, linguistic, socioeconomic, and demographic perspectives."[67]

4. *How was the "Model" Trained to Ensure it Respects and Promotes Democratic Values? Provide Examples.*

When we asked about how the models were trained to protect and respect democratic values, the four chatbots responded with different visions of democracy. For example, ChatGPT-4o responded that it uses RLHF "where diverse human annotators rated responses for helpfulness, truthfulness, and alignment with democratic values."[68] The bot also stated that it was "trained to not take sides in political debates and instead provide multiple perspectives while emphasizing fact-based reasoning."[69] GPT-4o gave an example comparing different economic and political systems. The chatbot stated, "Capitalism can drive innovation, while socialism emphasizes equity. Many countries use a mix of both." In contrast, Gemini 2.5 Flash responded to the prompt with examples of how it was trained to support democratic norms. First, the bot claimed that its training promoted inclusion of diverse political views through content from various political viewpoints, news sources, and cultural contexts."[70] Gemini also responded because it is also trained on information from Google Search and it can access real-time information from credible sources to combat misinformation. Lastly, Gemini stated that its developers utilized human feedback systems where "expert human reviewers, trained on guidelines that embody democratic values, provide feedback to the model."[71]

DeepSeek V3 responded to the prompt by stating that it was designed to "respect and promote democratic values—such as freedom of expression, equality, pluralism, and informed civic participation."[72] The system stated that it was trained on sources "that emphasize human rights, constitutional governance, and civic discourse, including legal documents, UN declarations, and balanced political analyses."[73] DeepSeek also stated, "While some argue authoritarian regimes can act faster, democracies ensure long-term stability, accountability, and protection of rights—key for sustainable development."[74] Finally, Grok 3, like Gemini stated that it was trained on a wide range of viewpoints, cultures, and ideologies.[75] However, also stated that it uses "consented user interactions on the X platform" while providing users control through opt-out mechanisms in its training. Finally, Grok described specific content moderation examples, such as refusing to generate content "that could incite unrest" and instead "explaining the importance of peaceful civic participation."[76]

5. *How Does User Feedback Affect the "Model" Development and Deployment to Align More with Responsible AI? Provide Examples.*

When asked how user feedback affects model development, all four bots provided examples of how they use such feedback to revise their models. But they did not then link these examples to responsible AI. GPT-4o responded that it used reinforcement learning as well as user feedback, especially when users flagged "responses as helpful, unhelpful, biased, inaccurate, or harmful."[77] The system provided specific product development examples, citing how "early users of GPT-4o in ChatGPT Plus reported issues with hallucinations in citations," leading OpenAI to add "a 'Double Check with Web' feature and improved citation formatting."[78] GPT-4o also responded to our prompt that feedback from "election integrity groups led to clearer limitations on GPT-4o providing voting information unless verified through trusted sources like CanIVote.org."[79]

Meanwhile, Gemini stated that users can "flag instances through 'thumbs down' feedback, report mechanisms, or more detailed qualitative feedback channels." The bot then described that this allowed developers to "refine training data: augment or re-balance datasets to ensure broader representation" and "develop bias mitigation techniques: implement new algorithms or fine-tuning methods to de-bias the model's outputs."[80] Gemini offered a real-world example, noting that user reports about "subtly misleading" medical advice would "trigger a review of the prompt and the model's response, leading to adjustments in the safety guidelines or model behavior."[81]

DeepSeek V3 responded that it analyzed user feedback to fine-tune its models using reinforcement learning. DeepSeek stated that when "users complain about the model being overconfident in wrong answers," developers implemented "more 'I don't know' or uncertainty-aware responses."[82]

Finally, Grok 3 stated that it relied on user feedback and the X platform "to refine the model's behavior."[83] Thus, when Grok 3 was criticized on X for generating responses that appeared to endorse controversial views (e.g., antisemitic content), user feedback prompted xAI to implement stricter moderation filters."[84]

Taken in sum, the researchers found the bots responded in a similar manner to our questions, although the responses varied in specificity. They all responded to our prompts, citing reinforcement learning, diverse data curation, bias detection algorithms, content moderation systems, and user feedback integration. They claimed to promote democratic values and refuse harmful requests but offered few concrete examples linking these to responsible AI. As a result, we had little understanding of how responsible AI principles guided training and in turn, affected their responses.

## Conclusion

After reviewing our three analysis strategies, we found that AI developers have not made responsible AI a priority. Google and, to a lesser extent, OpenAI discussed responsible AI on their websites. However, when we compared technical documents, we found these firms only occasionally referred to the terms of responsible AI. All four companies made significant mention of safety, and to a lesser extent, openness, privacy, and diversity. Hence, companies tend to focus more on mandated aspects of AI responsibility than non-mandated or soft-law issues such as accountability, explainability and/or interpretability.

When comparing responses to our questions, all four bots provided broad descriptions of why concerns such as human rights or democratic values were important. But when we prompted the bots for further details or specific examples, the bots provided few specifics. The response to our question about inclusiveness was an exception, where two bots responded with details about how they incorporated diverse ideas and voices.

Google appeared the most committed to responsible AI because it utilized both its technical documents and websites to signal the import of responsible AI practices. Google's chatbot also utilized more responsible AI terms when it responded to our prompts But in September,

DeepSeek signaled its commitment to openness and accountability by providing a wide range of technical information to make its LLM replicable. So, some analysts might assert its actions provide a strong commitment to responsible AI.

The researchers also conclude that the companies' commitment to responsible AI is not very deep. For example, all four bots said they were trained to protect privacy and to enable users to opt out of training if these users desired. However, recent reports have shown that both xAI and OpenAI used private conversations between users and the bots without direct permission. Moreover, these firms allowed such conversations to be web-scraped and searchable through standard search engines.[85]

In conclusion, although AI companies may claim to prioritize responsible AI, they don't appear to make it a top priority. Without clear guidelines for responsible AI, developers will struggle to act in a responsible manner. Moreover, society may struggle to hold developers accountable when the bots act in an inappropriate manner.

**Policy Implications**

Given our findings, policymakers should:

- Collaborate internationally to define responsible AI. AI is a global product built on global expertise and data, and officials must collaborate to ensure that AI benefits humanity.
- Building on that definition, examine whether transparency mechanisms such as the EU AI Act and Code of Practices might incentivize a stronger focus on responsible AI.
- Establish an international advisory council—comprising respected experts and citizen volunteers—to serve two key purposes: publicly calling out irresponsible or inconsistent practices by AI firms and recognizing exemplary responsible practices through an international award presented over a two-year period.